\scrollmode
\documentclass[sigconf]{acmart}
\usepackage{amsmath}
\usepackage{booktabs}
\usepackage{color}
\usepackage{array}
\usepackage{subfigure}
\usepackage{graphics}
\usepackage{threeparttable}
\usepackage{hyperref}
\usepackage{enumitem}
\usepackage{multirow}
\usepackage{stfloats}
\usepackage{courier}
\usepackage{hyperref}
\usepackage{bm}
\usepackage{makecell}
\usepackage{pifont}
\AtBeginDocument{%
  }

\setcopyright{acmcopyright}
\copyrightyear{`2018}
\acmYear{2018}
\acmDOI{XXXXXXX.XXXXXXX}

\acmConference[Conference acronym 'XX]{Make sure to enter the correct
  conference title from your rights confirmation emai}{June 03--05,
  2018}{Woodstock, NY}
\acmPrice{15.00}
\acmISBN{978-1-4503-XXXX-X/18/06}

\begin{document}

\title{Ensure Timeliness and Accuracy: A Novel Sliding Window Data Stream Paradigm for Live Streaming Recommendation}

\author{Fengqi Liang}
  \authornote{Equal contribution.}
\affiliation{%
  \institution{Kuaishou Technology}
  \city{Beijing}
  \country{China}
  }
\email{liangfengqi@kuaishou.com}

\author{Baigong Zheng}
  \authornotemark[1]
\affiliation{%
  \institution{Kuaishou Technology}
  \city{Beijing}
  \country{China}
  }
\email{zhengbaigong@kuaishou.com}

\author{Liqin Zhao}
  \authornotemark[1]
\affiliation{%
  \institution{Kuaishou Technology}
  \city{Beijing}
  \country{China}
  }
\email{zhaoliqin@kuaishou.com}

\author{Guorui Zhou}
  \authornotemark[1]
\affiliation{%
  \institution{Kuaishou Technology}
  \city{Beijing}
  \country{China}
  }
\email{zhouguorui@kuaishou.com}

\author{Qian Wang}
  \authornotemark[1]
\affiliation{%
  \institution{Kuaishou Technology}
  \city{Beijing}
  \country{China}
  }
\email{wangqian23@kuaishou.com}

\author{Yanan Niu}
  \authornotemark[1]
\affiliation{%
  \institution{Kuaishou Technology}
  \city{Beijing}
  \country{China}
  }
\email{niuyanan@kuaishou.com}





\begin{abstract}

Live streaming recommender system is specifically designed
to recommend real-time live streaming of interest to users.
Due to the dynamic changes
of live content, improving the timeliness of the live streaming recommender system is a critical problem.  Intuitively, the timeliness of the data determines the upper bound of the timeliness that models can learn. However, 
none of the previous works addresses the timeliness problem of the live streaming recommender system from the perspective of data stream design. 
Employing the conventional fixed window data stream paradigm introduces a trade-off dilemma between labeling accuracy and timeliness. In this paper, we propose a new data stream design paradigm, dubbed Sliver, that addresses the timeliness and accuracy problem of labels by reducing the window size and implementing a sliding window 
correspondingly. Meanwhile, we propose a time-sensitive re-reco strategy reducing the latency between request and impression to improve the timeliness of the recommendation service and features by periodically requesting the recommendation service. To demonstrate the effectiveness of our approach, we conduct offline experiments on a multi-task live streaming dataset with labeling timestamps collected from the Kuaishou live streaming platform. 
Experimental results demonstrate that  Sliver outperforms two fixed-window data streams with varying window sizes across all targets in four typical multi-task recommendation models. Furthermore, we deployed Sliver on the Kuaishou live streaming platform. Results of the online A/B test show a significant improvement in click-through rate (CTR),  and new follow number (NFN), further validating the effectiveness of Sliver.
\end{abstract}

\settopmatter{printacmref=true} 


\maketitle

\section{Introduction}
\begin{figure}[h] 

\includegraphics[height=4.8cm, keepaspectratio]{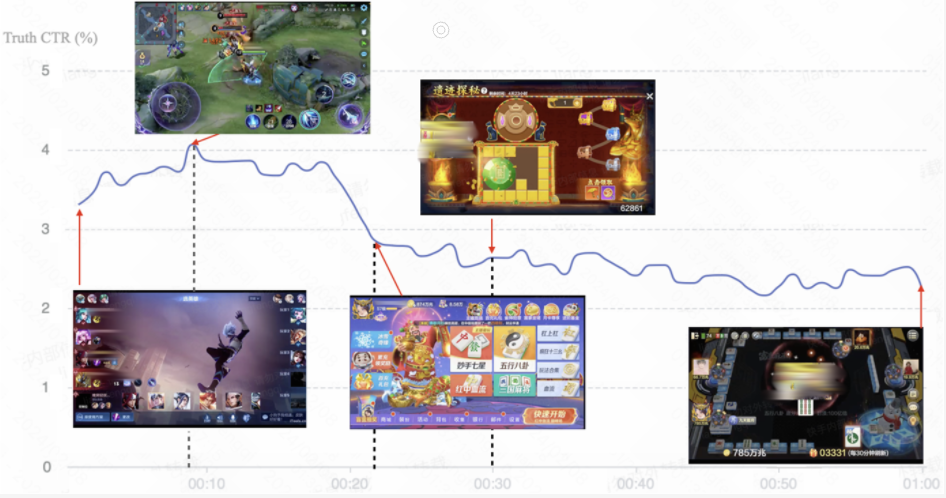}
	\caption{
This figure shows that the ground truth CTR undergoes dynamic changes in sync with the live content change in a live room. The game pictures in this figure represent the live content of the corresponding moment in the timeline. We can observe a significant change in live content and ground truth CTR between an hour ago and an hour later, as the anchor switched games. }

\label{fig:status}

\end{figure}

In the era of mobile internet, online live streaming has emerged as one of the most popular methods for online interaction and has experienced rapid development in recent years~\cite{deng2023contentctr}. Through live streaming platforms, anchors can share their experiences with audiences in real time. Various live streaming application scenarios have arisen, including online education~\cite{chen2021afraid}, e-commerce~\cite{yu2021leveraging}, etc.
Live streaming recommender system is specifically designed to recommend live streaming of interest to users on these live streaming platforms. Unlike traditional recommendation systems, live recommendations are uniquely characterized by the fact that the live content is constantly and dynamically changing ~\cite{rappaz2021recommendation} in real-time. For example, as shown in Figure ~\ref{fig:status}, dynamically changing live content leads to  the dynamic change of ground truth
Click-Through Rate (CTR) dynamic in a live room.  Consequently, ensuring the \textbf{timeliness} of live streaming poses a significant challenge for live streaming recommender system.

\begin{figure}[t]
\includegraphics[height=6.0cm, keepaspectratio]{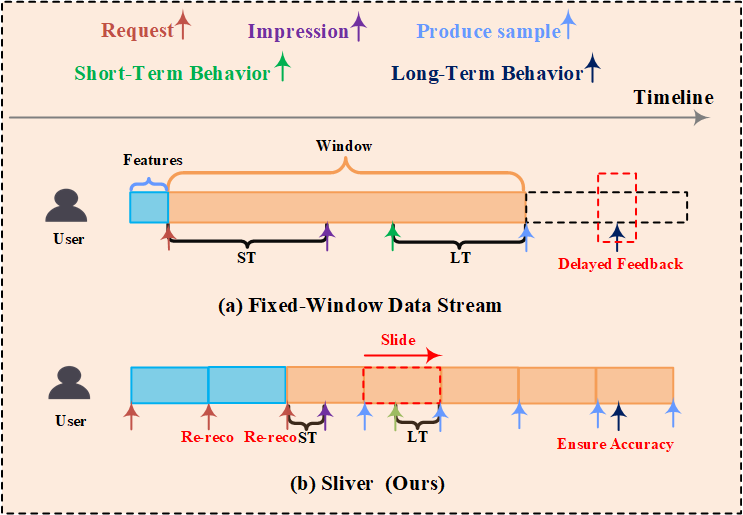}
	\caption{Comparison between conventional fixed-window data stream and our approach: conventional fixed-window data streams face a trade-off between the timeliness and accuracy of the live streaming recommender system. Sliver ensures both timeliness and accuracy through the implementation of sliding windows and re-reco strategy. ST and LT denote service timeliness and label timeliness respectively}\footnotetext{Source for image A}
	\label{fig:over}
\end{figure}

 Several works focus on addressing this problem from the model perspective and adding content information (e.g., time, multimodal information, or users’ dynamic behaviors) to the recommendation model to capture the dynamics of live streaming content~\cite{rappaz2021recommendation, zhang2021deep, gao2023live, deng2023contentctr}. However, all of these works overlook the crucial aspect of modeling timeliness in live streaming from the data perspective. Intuitively, the timeliness of the data determines the upper bound of the timeliness that models can learn. In other words, 
 a model may lack timeliness if it has not been trained with the latest samples.
In an industrial recommender system, data naturally presents itself in the form of a stream, and the recommendation model is incrementally updated as new data arrives (i.e., streaming learning)~\cite{chang2017streaming, chandramouli2011streamrec, wang2018streaming, wang2020streaming, wang2023streaming}. Streaming learning ensures a certain degree of timeliness in the live streaming recommender system because the online model can be updated immediately by the new live streaming samples. Nevertheless, one problem that streaming learning does not take into account is how to produce live streaming training samples and ensure 
 timeliness. For example, ideally, a user behavior (e.g., click) should become a training sample as soon as it occurs and then update the model by the streaming learning approach.  
 To address this problem, a straightforward
method is adopting the mainstream fixed time window data streams~\cite{moon2010online,ktena2019addressing,chen2022asymptotically} and shortening the time window size. However,
shortening the time window size will lead to a new problem of delayed
feedback ~\cite{joulani2013online, chapelle2014modeling, chen2022asymptotically}. For example, as shown in Figure ~\ref{fig:over} (a), the long-term behavior (e.g., follow) outside of the window is misclassified as a negative sample. Then a natural question could be asked: \textit{how to ensure the timeliness and accuracy of live data streams at the same time? 
}

 In this work, to the best of our knowledge, we first consider the question of how to design a data stream that ensures both timeliness and accuracy in live streaming recommender system.  To address this problem, we propose a novel \textbf{S}liding window data stream paradigm for \textbf{live} streaming \textbf{r}ecommendation (Sliver). As shown in Figure ~\ref{fig:over} (b),  Sliver slides a short time window and produces samples at the end of the sliding window. The latency between training samples and the real-world behaviors in the Sliver data stream is the length of the short time window, which inherently guarantees the timeliness of live-streaming recommender systems. 
Meanwhile, as we slide the window and know when the user exits the live room, we could leverage the explicit negative feedback behavior of exiting the live room as negative samples. This approach ensures the accuracy of negative samples and effectively addresses the problem of delayed feedback in fixed-window data streams. Finally, 
considering that online recommender systems are confined to using features and models only at the time of the request, a latency commonly exists between the request moment and impression moment. This delay detrimentally impacts the timeliness of recommendation features and services. To handle this challenge, we propose a time-sensitive strategy, dubbed re-reco, consistently \textbf{re-}requesting the \textbf{reco}mmendation service,  when the recommended live broadcast remains unimpressed to user. This approach ensures the ongoing timeliness of the features and the online recommendation service.

 To demonstrate the effectiveness of our data stream, we propose a real-world live streaming dataset with detailed time information collected from Kuaishou APP\footnote{\url{https://www.kuaishou.com/}}, one of the largest short-video and live-streaming platforms in China. Based on this dataset, we show the evolution of three different data streams of the live-streaming recommender system on Kuaishou platform, with latency ranging from hour level to minute level to second level, and timeliness gradually increasing. The first two are fixed-window data streams and the last one is our proposed Sliver data stream.
 We conduct offline experiments with these data streams based on the multi-task learning problem ~\cite{ma2018entire,ma2018modeling,tang2020progressive,chang2023pepnet} of predicting 
  multiple user interactions for live streaming with different targets (i.e., click, follow and like), which could capture
various behavior preferences of users. The click and follow targets are predicted in the impression space, and the like target is predicted in the post-click space~\cite{ma2018entire}.  
  Our experimental results show that our proposed Sliver data stream achieved state-of-the-art (SOTA) performance compared to two fixed window data streams on all targets, which illustrates the necessity to consider how to ensure both the timeliness and labeling accuracy of live recommendation from the perspective of the data stream. Finally, our proposed data stream is deployed on the Kuaishou APP and achieves a 6.765\%-8.304\% improvement of CTR, and 2.788\%-3.697\% improvement of new follow number (NFN) on two pages which further validates the effectiveness of our method.

 To summarize, this work makes the following contributions:
\begin{itemize}[leftmargin=*]
\item To the best of our knowledge, we first consider how to ensure the timeliness and accuracy of live streaming recommender system at the same time from the data stream perspective.  

\item We propose a novel Sliver data stream paradigm, sliding a short time window and producing samples
at the end of this sliding window continuously, which inherently guarantees timeliness. The sliding window manner allows us to adopt the explicit negative feedback of exiting the live
room to avoid delayed feedback and ensure label accuracy. 

\item  We propose a time-sensitive re-reco strategy to ensure the timeliness of recommendation features and whole recommendation services.

\item We open source a real-world live-streaming dataset with labeling timestamp information and conduct offline experiments. Moreover, we deploy Sliver data stream on  Kuaishou live streaming platform and conduct online experiments. Experiment results show that our proposed data stream outperforms previous data streams which have higher latency and less accuracy in both offline and online environments. 
\end{itemize}


\section{Related Works}
\subsection{Live Streaming Recommendation}
With the development of live streaming as an emerging social media,  increasing attention has been
given to live streaming recommendations. Unlike the traditional problem of recommending static items ~\cite{he2014practical, wang2018billion,zhou2019deep,zhou2019deep,pi2020search}, the content in live streaming is dynamically changed.  Several works has been proposed to handle live live-streaming recommendation problem. 
LiveRec ~\cite{rappaz2021recommendation} add time content to self-attentive mechanisms for historical interactions and make models more time-sensitive.
 ~\cite{zhang2021deep} utilizes LSTM and attention-based models to extract both streamers' and audiences' preferences for recommendation. 
DRIVER~\cite{gao2023live} learns dynamic representations by leveraging users’ highly dynamic behavior in the live room. ContentCTR~\cite{deng2023contentctr} adopts the multimodal transformer to extract multimodal information in live streaming frames.   All of these works focus on the model
perspective to handle timeliness problems in live streaming recommendations. However, our work focuses on more crucial aspects of
modeling timeliness in live streaming from the data perspective.

\subsection{Streaming Recommendation and Delayed Feedback}
To address real-world dynamics such as continuous shifts in user preferences, streaming recommendation~\cite{chang2017streaming, chen2013terec, devooght2015dynamic, wang2018streaming, he2023dynamically} is suggested. This approach involves the dynamic and simultaneous updates of both data and recommendation models along the timeline. Likewise, streaming recommendation could handle the problem of dynamic content change in live streaming recommendation. However, all of this work does not consider how to design a data stream to meet the timeliness requirement of live streaming recommendations. Shortening the time window size in the fixed window paradigm will lead to a new problem of delayed feedback~\cite{joulani2013online, chapelle2014modeling, chen2022asymptotically}. To address the problem of delayed feedback, several works~\cite{chapelle2014modeling, yoshikawa2018nonparametric} propose to model the delay time distribution. However, these works only attempted to optimize
the observed behavior information rather than the actual delayed
behavior, which cannot fully utilize the positive feedback.  Another mainstream approach~\cite{ktena2019addressing,lee2012estimating,gu2021real,chen2022asymptotically} to addressing delayed feedback  
employs sample duplicating mechanisms to reuse delayed behavior outside the fixed observation window and adopts importance sampling to de-bias the data distribution, However, these approaches face the problem of the trade-off between timeliness and label accuracy, especially for live recommendations, a more time-sensitive problem. Our approach could ensure both labeling accuracy and timeliness through sliding window paradigm.  

\subsection{Multi-Task Learning for Recommendation}  
Multi-task learning~\cite{caruana1997multitask,ruder2017overview} is a learning paradigm in machine learning that aims to leverage valuable information from multiple related tasks to enhance the generalization performance of all the tasks. 
A modern industrial recommendation model employs a multi-task learning approach to capture diverse user behavior preferences. It predicts the combined value of recommended items by blending targets derived from different behaviors. 
The most common multi-task learning is
Shared Bottom ~\cite{ruder2017overview}, which employs hard parameter-sharing to predict each target individually based on the shared output.
To deal with the differences between
tasks, MMOE~\cite{ma2018modeling} shares all experts like MOE~\cite{jacobs1991adaptive} and further utilizes
different gating networks to obtain different fusion expert weights for each task.
ESSM ~\cite{ma2018entire} relies on a soft parameter-sharing structure and concurrently optimizes two interrelated tasks using sequential modes to address the sparsity issue in the prediction target.  To deal with task conflicts and alleviate negative transfer
phenomenon, CGC and PLE ~\cite{tang2020progressive} set up shared and task-specific experts for
each task. 
The aforementioned multi-task learning models complement our approach, and our proposed data streaming can be adopted as a plug-in to augment the online performance of multi-task models in live streaming recommender systems.

\section{METHODOLOGY}
In this section, we start by formalizing the multi-task live streaming recommendation problem in streaming data settings and explaining the key notations. Afterward, we elaborate on the proposed Sliver data stream and the evolution of data streams 
in Kuaishou live streaming recommendation scenario. Finally, we show the modeling approach with the proposed Sliver data stream. 

 \begin{figure}[h]
	\includegraphics[height=2.6cm, keepaspectratio]{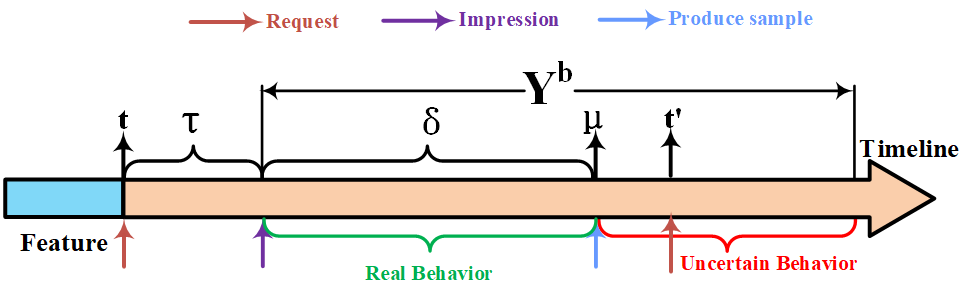}
	\caption{An illustration of the timeline for producing a sample in the streaming recommender system. The time interval between request, impression, and sample production is $\tau$ and $\delta$ correspondingly. $Y^{b}$ is a random variable for the moment of the user behavior that happens in the real world after the impression.  
                }
	\label{fig:sl}
\end{figure}

\subsection{Problem  Definition}
   We begin by defining the multi-task live streaming recommendation problem in streaming data settings. Let $\left\{D_\mu \right\}_{\mu=1}^{\infty}$ denote the data stream, where $D_\mu$ denotes the live training samples at specific timestamp $\mu$.  $D_\mu = \left\{
   U_{t}, I_{t}, A_{t}, Y^{b}_{\mu} = f_{\mu}(y_{b}) \right\} $, where $t$ is the moment of requesting recommendation service, $U_{t}$, $I_{t}$, $A_{t}$ and $Y^{b}_{\mu}$ denote the users, live broadcasts, anchors and collected user behaviors $b$ (e.g., click, like and follow) as labels correspondingly. As shown in Figure~\ref{fig:sl}, the recommended live broadcasts are impressed to users after $\tau$ time. Finally, after $\eta$ time at the moment $\mu$, training samples are produced and the online recommender model is incrementally trained by these samples. Let $Y^{b}$ be the random variable and its value $y^{b}$ represents the moment that the user behavior (e.g., click) happens in the real world after the impression. If  $y^{b} < \mu$, $f_{\mu}^{b}(y^{b}) = 1$ will act as a positive feedback. If $y^{b} > \mu$,  $f_{\mu}^{b}(y^{b})$ will be uncertain feedback at the moment $\mu$ since we don't know whether the behavior will eventually happen or not at this moment. 
   Our goals are to estimate the probability of multiple user behaviors $P({{Y_{t^{'}}}^{b}=1 \mid U_{t^{'}}, I_{t^{'}}, \theta_{\mu}})$ at the moment $t^{'}$ of the next user requests, where $\theta_{\mu}$ is the parameter of the multi-task learning model incrementally learned through $D_\mu$. However, as mentioned in the introduction,  the live broadcast features $I$ constantly and dynamically change over time in the live streaming recommendation task. In other words, there is a distribution shift of $I$ at moments $t$ and $t^{'}$ which leads to the distribution shift of user behavior. Meanwhile, the recommendation results may become outdated after a latency period $\tau$, even if the initially recommended results were timely at the request moment $t$. Thus the time delay $\tau$ decides the timeliness of \textbf{the whole recommendation service}.
    On the other line, the positive label $Y_{\mu}^{b}$ may have a delay of up to $\delta$ if user behaviors occur directly at the moment of $t+\tau$. In summary, the timeliness of training samples and recommendation service is related to both time delays $\tau$ and $\delta$. \textbf{Note} that the model parameters $\theta_{\mu}$ used for inference are learned based on the delayed training samples and the higher the timeliness of the training samples, the more timely results which the model can provide. This is the motivation behind our exploration of how to address the timeliness of live recommendations in terms of the sample production and data stream perspective.

\subsection{The Evolution of Live  Data Stream in Kuaishou APP}
\label{sec:3.2}
\begin{figure*}[t]
	\includegraphics[width=\textwidth]{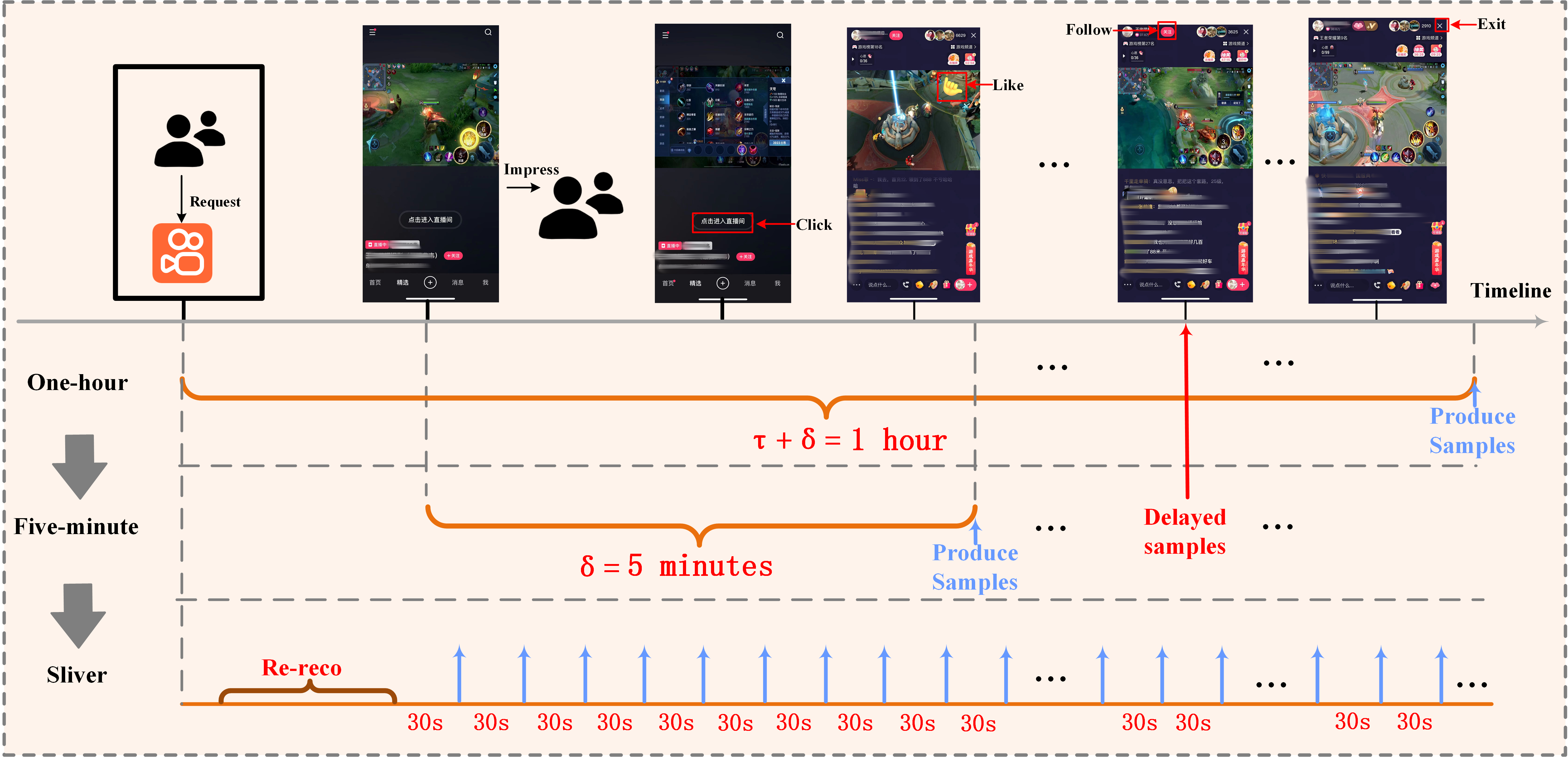}
	\caption{The overview data streaming evolution in Kuaishou live streaming platform from one-hour data stream to five-minute data stream to Sliver data stream with increasing timeliness. One-hour data stream adopts the moment of request as the starting point for the one-hour window. Five-minute data stream adopts the moment of impression as the starting point for the one-hour window. Our proposed Sliver data streaming utilizes 30s sliding window to balance timeliness and accuracy.}
	\label{fig:dsc}
\end{figure*}

Figure~\ref{fig:dsc}  illustrates the evolution of live data streams in the Kuaishou APP. We begin with two fixed-window data streams, with a window size of one hour or five minutes, respectively. Then we present our proposed Sliver data stream which produces samples in thirty-second intervals. 

\subsubsection{Fixed-Window Data Stream Paradigm}
A mainstream data stream paradigm~\cite{moon2010online,ktena2019addressing,chen2022asymptotically} for an industrial recommendation system is to produce samples involves initiating the process at a specific time (e.g., user requests and item impression), recording user behavior throughout a predefined time window \textbf{$\textbf{w}$}, and then producing samples at the end of that window. Following this idea, we constructed the first data stream of the kuaishou live recommender system with \textbf{one hour} time window $w_{h}$ starting from the request time. This can be formalized as:
\begin{equation}
\begin{aligned}
\tau + \delta  = w_{h}. 
\end{aligned}
\end{equation}
As shown in the top of Figure ~\ref{fig:dsc}, if the user requests the recommendation service at the moment \textbf{$t$}, then the user behaviors $b$  in  $t < y^{b} < t + w_{h}$ will be recorded and take as the positive training samples for online learning at the moment $\mu =  t+w_{h}$. On the contrary, if the recommended live broadcast has impressed but user behavior does not occur in the window, this sample will be adopted as negative feedback. This process can be formalized as:
\begin{equation}
\begin{aligned}
f_{\mu}^{b}(y^{b})= \begin{cases}1,  &  t< y^{b} <  t+w_{h} \\ 0, &   y^{b} >  t+w_{h} \  \& \ \tau <  w_{h} \ \& \ b \in S_{imp}
\\ 0, &  y^{b} >  t+w_{h} \ \& \ t< \eta_{c} <  t+w_{h} \ \& \  b \in S_{post} 
\end{cases},
\end{aligned}
\end{equation}
where $S_{imp}$ is the impression space, $S_{post}$ is the post-click space, and $\eta_{c}$ is the click moment. The click and follow behaviors are in the impression space and the like behavior is in the post-click space.

However, in the context of live broadcasts, there is a significant disparity in the data distribution of live content an hour before and an hour after the broadcast. This incongruity poses a considerable challenge to the timeliness of live recommendation systems. Meanwhile, there is a few-minute time delay $\tau$ between the moment of request and the moment of exposure 
 which also affects the timeliness of the whole recommendation service.
Second,  if a user's follow behavior occurs midway through the one-hour window, it takes an additional half-hour delay for this positive feedback to be incorporated as a training sample. Finally, in this data stream, the time delay $\tau + \eta$ of the live broadcast feature is a whole hour which lacks the timeliness of the live broadcast features. 
\begin{figure}[t] 
\includegraphics[height=5.5cm, keepaspectratio]{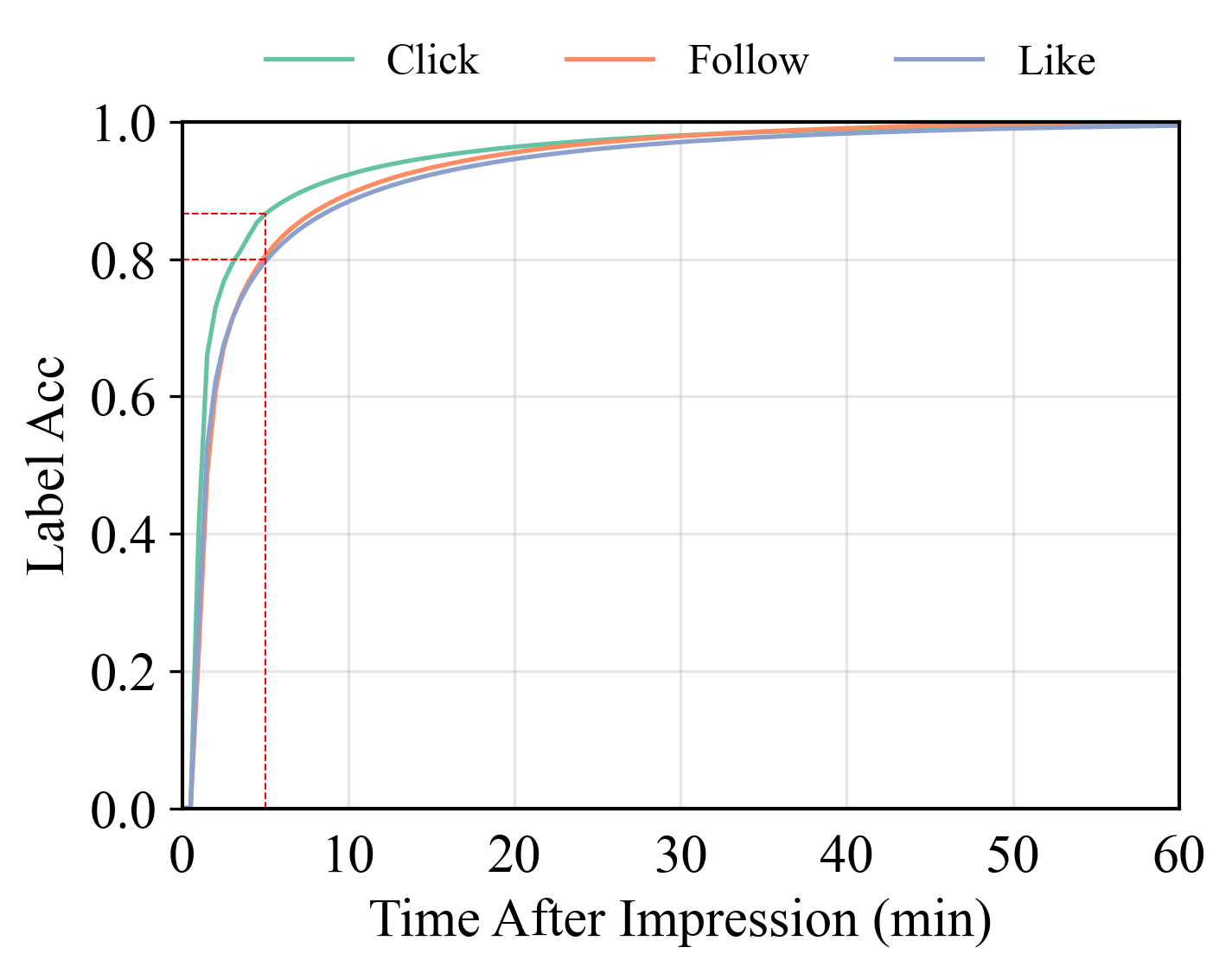}
 \caption{
 This figure shows the change in labeling accuracy over time in the Kuaishou live streaming platform. Five minutes after impression, the click label accuracy is about 86\%, and the like and follow label accuracy is about 80\%.
}
 \label{fig:acc}
\end{figure}

To alleviate this problem, as shown in the middle of Figure~\ref{fig:dsc},  we reduce the time window to \textbf{five minutes} and start from the live broadcast impression with the time window $w_{m}$.  
\begin{equation}
\begin{aligned}
\delta  = w_{m}. 
\end{aligned}
\end{equation}
The reason we don't consider the request time as a starting point is that the latency between the request time and the impression time may be more than five minutes which leads to most of the unimpressed samples being treated as negative samples.  The label logic of five minutes data stream can be formalized as:
\begin{equation}
\begin{aligned}
f_{\mu}^{b}(y^{b})= \begin{cases}1,  &  t + \tau < y^{b} <  t + \tau+w_{m} \\ 0, &   y^{b} >  \mu \ \& \ b \in S_{imp}
\\ 0, &   y^{b} >  t+\tau+w_{m} \ \& \ t + \tau < \eta_{c} <  t + \tau+w_{m} \ \& \ b \in S_{post}
\end{cases}.
\end{aligned}
\end{equation}

This approach shortens the latency of $\delta$ and somewhat improves the live samples' timeliness.
However, reducing the time window size will create a  new problem of delayed feedback~\cite{joulani2013online, chapelle2014modeling, chen2022asymptotically}. Take Figure~\ref{fig:dsc} as an example: If a user's follow behavior occurs outside of the five-minute window, this sample will be taken as a fake negative sample at training time due
to delayed feedback. As shown in Figure~\ref{fig:acc}, the label accuracy at the moment five minutes after the impression is about 86\% for click and 80\% for follow and like. This problem is not serious in the one-hour data stream, because the labeling accuracy increases as the window size increases~\cite{chen2022asymptotically}.  
Meanwhile, the same time delay $\tau$ with the one-hour time window data stream also affects the timeliness of the live broadcast feature and the whole recommendation service. 

 In summary, training samples under fixed-window data streams face the following two timeliness problems:
\begin{enumerate}
\item Fixed-window data streams inevitably involve a trade-off between label timeliness and label accuracy.   

\item Latency $\tau$ between request and exposure limits the timeliness of features and the whole recommendation service.  
\end{enumerate}

  We address the first problem by proposing the new sliding window data stream paradigm and handle the second problem by the re-reco strategy.
  
 \subsubsection{Sliding Window Data Stream Paradigm} 
 \label{sec:2}
 The fundamental reason for the trade-off between labeling accuracy and timeliness in the fixed-window data stream is that $y^{b}$ is uncertain and produces samples at the end of window truncated real-world distribution $Y^{b}$. Inspired by the idea of calculus, if we treat the time window as a differential, then by constantly sliding the window to do integral, we approximate the real-world distribution $Y^{b}$. The differential ensures timeliness and the integral ensures accuracy.  This idea motivates us to propose the sliding window data stream paradigm.
 
 As shown in the bottom of Figure ~\ref{fig:dsc}, our proposed Sliver adopts a uniform time $t_{uni}$ as the starting point and produces samples at the end of each sliding window. Then we assign a uniform ID index for each sliding window. Then the producing sample moment $\mu_{k}$ can be represented as follows:

\begin{equation}
\begin{aligned}
\mu_{k} = t_{uni} + k * w_{s},
\end{aligned}
\end{equation}
where  $k$ is the window ID and $w_{s}$ is the window size. We choose 30s as $w_{s}$ in the practical application. This approach shortens the delay $\delta$ to 30s and ensures both label timeliness and accuracy. 
 Next, we describe the label logic in the Sliver data stream which can be formulated as follows:
 \begin{equation}
\begin{aligned}
f_{\mu_{k}}^{b}(y^{b})= \begin{cases}1,  &  \mu_{k-1}< y^{b} < \mu_{k} \\ 0, &  y^{b} > \mu_{k} \ \& \ \mu_{k-1}<\eta_{exit}<\mu_{k} \ \& \ b \in S_{imp} \\
0, &  \eta_{click} < \mu_{k}  \ \& \  y^{b} > \mu_{k} \ \& \ \mu_{k-1}<\eta_{exit}<\mu_{k} \ \& \ b \in S_{post}
\end{cases},
\end{aligned}
\end{equation}
where $\eta_{exit}$ is the moment of user exits the live broadcast.
 
In summary, the proposed sliding window  data stream has the following benefits compared to fixed time windows:
\begin{enumerate}
\item \textbf{Low latency:} The latency $\delta$ from user behavior to sample generation is less than 30 seconds. Thus, it is more time-sensitive for the live recommendation task with dynamically changing live content.   

\item \textbf{High accuracy:} Our approach adopts the fact that the viewer did not perform the behavior and exited the live room as the negative feedback which is an accurate negative label.

\end{enumerate}

 \subsection{Modeling Under Slide Window Data Stream}

\subsubsection{Multi-task  Model in Ranking Phase}
The varied user behaviors occurring concurrently within a live broadcast may contribute to the effectiveness of recommendations. Consequently, we employ a unified multi-task model to capture the shared characteristics of diverse user behaviors occurring simultaneously. The input of the multi-task model can be formally defined as follows:
 \begin{equation}
\begin{aligned}
x^{\mu}=[x_{i}^{t}, x_{u}^{t}, x_{a}^{t}].
\end{aligned}
\end{equation}
where  $x_{u}$, $x_{i}$, and  $x_{a}$ are the user features, live broadcast features and anchor features correspondingly. The prediction at moment $m$ for task $b$ is
formulated as:
 \begin{equation}
\begin{aligned}
\hat{y}^{b}_{\mu}=f(x^{\mu}, \theta_{s}, \theta_{b}),
\end{aligned}
\end{equation}
where $\theta_{s}$ and $\theta_{b}$ denote the  
task-shared and task-unique parameters for multi-task model $f$ respectively. We use a multi-objective loss to optimize our model:
\begin{equation}
L= \sum_{b\in B} w^b*L^b,
\end{equation}
where $w^b$ and $L^b$ are the weight and loss for each task $b$.
We adopt standard log-likelihood function~\cite{zhou2018deep} to optimize different task targets.
\begin{equation}
L^{b}_{\mu}=-\frac{1}{N^{b}} \sum(y_{\mu}^{b} \log \hat{y}_{\mu}^{b}+(1-y_{\mu}^{b}) \log (1-\hat{y}_{\mu}^{b})),
\end{equation}
where $N^{b}$ is number of samples for behavior $b$ at the moment $\mu$. Finally, we use the multi-objective prediction results to compute a fusion score $s_{t+1}$ for online recommendation ranking at the moment $t+1$:
\begin{equation}
s_{t+1}=\sum_{b \in B}  \alpha^b*\hat{y}^b_{t+1},
\end{equation}
where $\alpha^b$ is the fusion weight for each target.
\subsubsection{A Time-Sensitive Re-reco Strategy}
\label{sec:3}
\begin{figure}[h]
	\includegraphics[height=6.5cm, keepaspectratio]{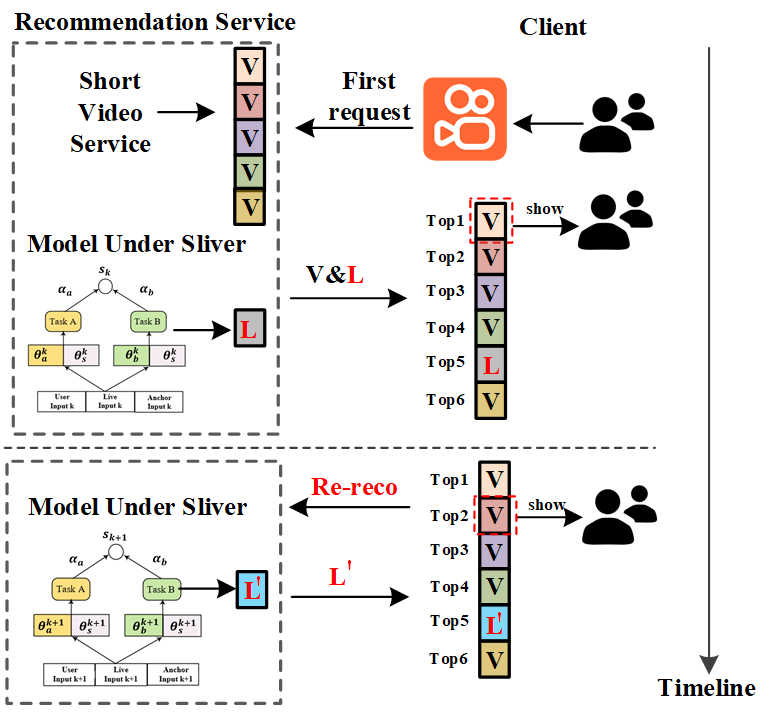}
	\caption{An illustration of the time-sensitive re-reco strategy. We recursively call the live recommendation model before the live impression.}
	\label{fig:rereco}

\end{figure}

Although the model learned under the Sliver data stream is time-sensitive, the entire recommendation live result still lacks timeliness because there is a non-trivial latency between the client requesting the live recommendation service and the actual impression of the live broadcast. As shown in Figure~\ref{fig:rereco}, the main reason leading to this time delay in our live streaming scenario is that the recommended live broadcast is mixed with recommended short videos in one time request and it is uncertain when the recommended live broadcast is impressed to user.   To address this problem, we re-request the time-sensitive live recommendation model (re-reco) every thirty seconds when the live room is unimpressed then use the re-requested live broadcast recommendation result to replace the original recommendation. This strategy shortens the time delay $\tau$ and ensures the timeliness of the feature. Meanwhile, this strategy ensures the timeliness of the whole online recommendation service. 




\section{EXPERIMENTS}
In this section, we conduct extensive experiments to demonstrate
the effectiveness of our proposed data streaming. First, we evaluate the
proposed methods in offline settings, and then report the online
experimental results on the Kuaishou App. Finally, we present
how the proposed sliver data stream is deloyed in
Kuaishou live streaming platform.

\subsection{Offline Evaluation}
\subsubsection{Dataset}
\begin{table}[h]
\small
\renewcommand\arraystretch{1.3}
\tabcolsep=0.3cm
   \caption{Dataset Information}
    \label{tab:fea_table} 
        \begin{tabular}{c|c}
          \toprule
          \textbf{}  & \textbf{Type} \\
          \midrule
          User Features & ID, Gender, Age, City, Click Anchor History \\
          Live Features & ID,  Live type \\
          Anchor Features & ID, Gender, Anchor type \\
          Behaviors & Impression, Click, Follow, Like, Exit \\ 
          \bottomrule
        \end{tabular}
	\end{table}
Due to a scarcity of research and datasets on the temporal dynamics of live-streaming recommendation systems, we collect and open source an industrial live dataset with detailed timestamp information from Kuaishou platform,
which boasts over three million daily active live streaming users. We extract a subset of three day logs (according to the request timestamp) from Dec. 27th to Dec.
29th, 2023  by sampling 1\% of users. As shown in Table~\ref{tab:fea_table}, for business privacy reasons, our dataset contains only a representative portion of the features and behaviors of our online recommendation system.  For each behavior, we provide information on the exact timestamp when the behavior occurred.   According to the timestamp information,  we split three data streams as mentioned in section ~\ref{sec:3.2}.  

\begin{table*}[t]
\small
\renewcommand\arraystretch{1.3}

   \caption{Result of different data streams on offline Kuaishou live streaming dataset in four typical multi-task models. The notation $\uparrow$ and $\downarrow$ denote RelaImpr of average AUC.
}
    \label{tab:main_table} 
        \begin{tabular}{c|ccc|ccc|ccc}
          \toprule
          \multirow{2}{*}\textbf{Data}  & \multicolumn{3}{c|}{\textbf{One-Hour Window (base)}} &  \multicolumn{3}{c|}{\textbf{Five-Minute Window}} & \multicolumn{3}{c}{\textbf{Sliver}} \\
          \cmidrule{2-10}
          {} & \textbf{Click} &  \textbf{Follow} & \textbf{Like}  & \textbf{Click}  & \textbf{Follow} & \textbf{Like} & \textbf{Click}  & \textbf{Follow} & \textbf{Like}  \\
          \midrule[0.8pt]
          Shared Bottom & 0.6724  &  0.6621  & 0.6839  &  0.6779 $\uparrow_{3.19\%}$   & 0.6656 $\uparrow_{2.16\% }$ & 0.6808 $\downarrow_{1.69\%}$ & \textbf{0.6811}  $\uparrow_{5.04\%}$ & \textbf{0.6690} $\uparrow_{4.26\%}$ & \textbf{0.6924} $\uparrow_{4.62\%}$ \\
          MMOE & 0.6757  & 0.6613 & 0.6817  &  0.6798 $\uparrow_{2.33\% }$  & 0.6635  $\uparrow_{1.36\% }$ & 0.6842  $\uparrow_{1.38\% }$ &\textbf{0.6837} $\uparrow_{4.55\%}$ & \textbf{0.6714} $\uparrow_{6.26\%}$ & \textbf{0.6959} $\uparrow_{7.82\%}$\\   
          CGC  & 0.6764  & 0.6646& 0.6843  & 0.6822  $\uparrow_{3.29\% }$ & 0.6654  $\uparrow_{0.49\% }$  & 0.6869  $\uparrow_{1.41\% }$ &  
          \textbf{0.6855} $\uparrow_{5.16\%}$ &  \textbf{0.6722} $\uparrow_{4.61\%}$ & \textbf{0.6972} $\uparrow_{7.00\%}$\\ 
          PLE &  0.6740 & 0.6624  &  0.6804  &      0.6791  $\uparrow_{2.93\% }$ & 0.6619 $\downarrow_{0.31\%}$  & 0.6831  $\uparrow_{1.50\%}$& \textbf{0.6837} $\uparrow_{5.57\%}$ & \textbf{0.6686} $\uparrow_{3.81\%}$ & \textbf{0.6917}$\uparrow_{6.26\%}$  \\ 
          \bottomrule
        \end{tabular}
        \label{tab:main}
	\end{table*}

\subsubsection{Metric}
We adopt the data before 7:00 p.m. Dec. 29th, as training data. To simulate online scenarios, for one-hour data stream, 
the final request timestamp of the training sample is 6:00 p.m.; For five-minute data stream, the final impression timestamp of the training sample is 6:55 p.m. Following ~\cite{wang2023streaming},  we adopt the five hour data after 7:00 p.m. as test data and incrementally evaluate the trained models under streaming learning settings. For example, after evaluating the data between 7:00 p.m. and 8:00 p.m., we increase the training data by one hour and evaluate the results for the next hour between 8:00 p.m. and 9:00 p.m.. To align real-world scenarios, we used the absence of positive behavior and exit the live room as the negative sample. We apply the average AUC  (five times)~\cite{cheng2016wide} to evaluate the models trained under different data streams and report
the average AUC score for each task individually.   Besides, we adopt RelaImpr metric to measure
relative improvement over different data streams. The RelaImpr is defined as follow:
$$
\text { RelaImpr }=\left(\frac{\text { AUC }(\text { measured model })-0.5}{\text { AUC }(\text { base model })-0.5}-1\right) \times 100 \%
$$

\subsubsection{Baselines.}
To illustrate the effectiveness of our proposed data stream paradigm, 
we compare it with two fixed window baselines on typical multi-task recommendation models, including: 
\begin{itemize}[leftmargin=*]
\item \textbf{Shared Bottom}~\cite{ruder2017overview}  shares
the parameters of the bottom DNN layers and employ task-specific
towers to generate corresponding scores. We adopt a two layer bottom DNN with hidden size (64,32).
\item \textbf{MMoE}~\cite{ma2018modeling} shares several experts and a  task-specific gating network across all tasks to model relationships between different tasks. In our implementation, we set three experts with hidden size (64, 32).

\item \textbf{CGC}~\cite{tang2020progressive} adopts independent experts for each task and retains the shared experts for all tasks. We set the number of task-specific experts and shared experts to 1 and all the experts with hidden size (64, 32).  

\item \textbf{PLE}~\cite{tang2020progressive} is the version of multi-layer CGC model. In our implementation, we use two CGC layers and set the hidden size of the first CGC layer (64, 32) and the hidden size of the second CGC layer (32, 32). 
\end{itemize}
For the task tower in all models, we adopt a three-layer DNN with hidden size (32, 32, 16).

\subsubsection{Implementation Details}
We implement all the models based on TensorFlow~\cite{abadi2016tensorflow}. We use Adam ~\cite{kingma2014adam} for
optimization with 0.001 learning rate. The batch size
is set as 4096 in offline training and the embedding size for the ID feature and side feature is fixed to 32 and 8 for all models correspondingly and ignore user ID in offline training to ensure result stability.
We adopt xavier initialization~\cite{glorot2010understanding} to initialize the parameters.

\begin{table}[t]
   \caption{Online A/B tests of upgrading one-hour data stream to five-minute data stream}
    \label{tab:up1} 
        \begin{tabular}{c|cc}
          \toprule
          \textbf{}  & Featured Page & Single-Columned Page \\
          \midrule
          CTR &  +13.653\% & +9.470\% \\
          NFN  & +6.091\% & +5.666\% \\
          \bottomrule
        \end{tabular}
\end{table}

\begin{table}[t]
   \caption{Online A/B tests of upgrading five-minute data stream to Sliver data stream}
    \label{tab:up2} 
        \begin{tabular}{c|cc}
          \toprule
          \textbf{}  & Featured Page & Single-Columned Page \\
          \midrule
          CTR &  +8.304\% & +6.765\% \\ 
          NFN  & +3.697\% & +2.788\% \\
          \bottomrule
        \end{tabular}
\end{table}

\subsubsection{Results Analysis.}
The results of four typical multi-task models on three tasks under different data stream settings are shown in Table ~\ref{tab:main}. From the table, we can observe that our proposed Sliver data stream outperforms two-fix window data stream on all models and targets.   Specifically, Sliver get 4.55\%-5.57\% RelaImpr on average AUC improvement on click behavior, 3.81\%-6.26\% on follow behavior, and 4.62\%-7.82\% on like behavior, which demonstrates the effectiveness of Sliver to ensure both timeliness and accuracy in live streaming recommendation task.

 Meanwhile, we can observe that the most click performance improvement comes from  improvement of timeliness and simply reducing the time window from one hour to five minutes can get 2.33\%-3.29\% RelaImpr and adopt reducing the window size to Sliver will further improve the performance for click target. However, in follow and like targets, reducing fixed window size may not result in performance improvements in some cases which illustrates that label accuracy is also important. Our Sliver data stream achieves the best performance in all cases by ensuring both the timeliness and accuracy of labels.
   
   On the other line. we can also find that the CGC model gets the best performance compared with the Shared Bottom model and MMOE model which align the conclusions with ~\cite{tang2020progressive}. PLE doesn't perform as well as a single layer of CGC. A possible reason is that the multi-layered structure may not suit our scenario.

\begin{figure*}[t]
	\includegraphics[height=5.3cm, keepaspectratio]{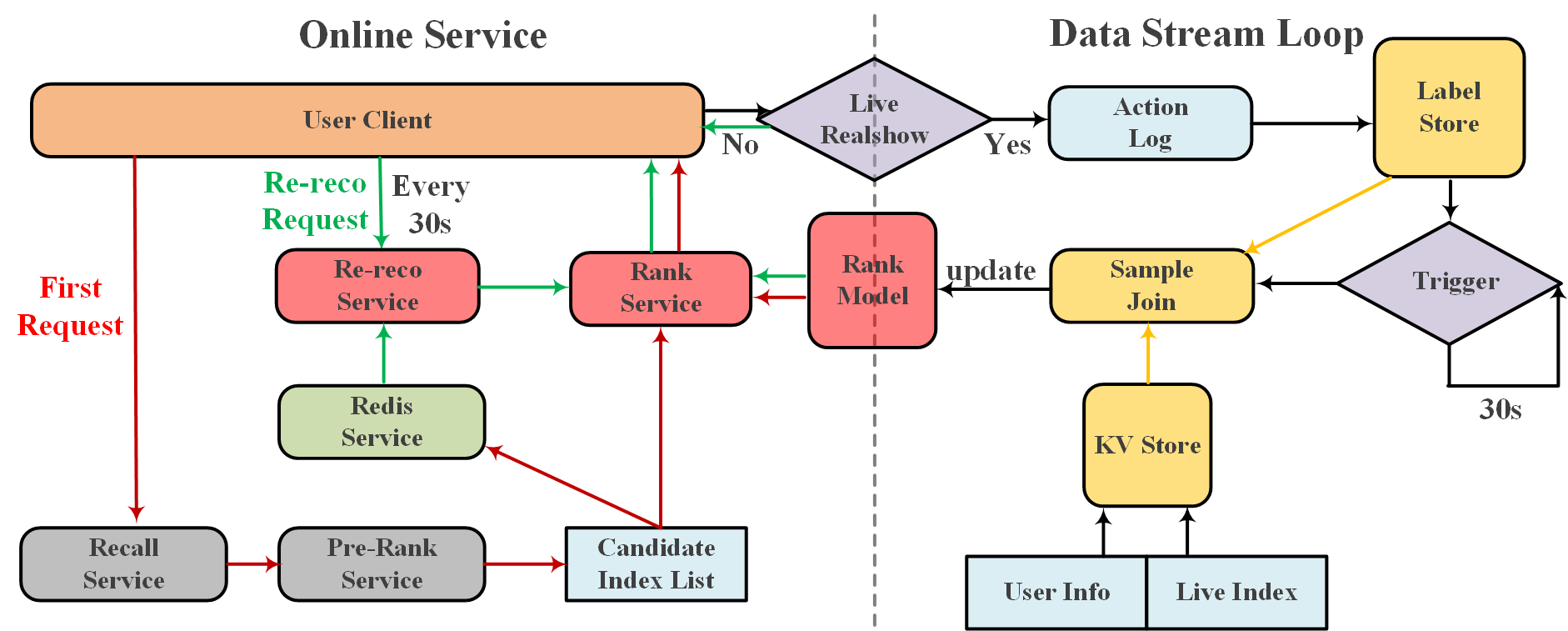}
 \caption{This figure shows the architecture of the recommender system on Kuaishou live streaming platform. The left part
shows the workflow of the online service including the first request service and the re-reco service.  The right part shows the workflow of our proposed sliver data stream.}
	\label{fig:service}
\end{figure*}

\subsection{ Online A/B Test}
We deploy our proposed Sliver data stream in the Kuaishou app which is one of the largest live-streaming platforms in China and conduct the online A/B testing on the featured page and single-columned page. For the detailed page information, please refer to Appendix~\ref{app:A}. We adopt the metric of CTR and new follow number (NFN) to evaluate the online data stream. NFN is an important metric in live streaming products because viewers like to watch their favorite anchors repeatedly and these anchors can be found explicitly by the follow relationship. As our online data stream goes through two upgrades, transitioning from one-hour to five-minute and eventually to Sliver, we present our online results in two separate parts.

Table ~\ref{tab:up1} shows the improvements of five-minute data streaming the online A/B test for four days compared to one-hour data streaming. We can find that five-minute data streaming achieves average improvements of  13.653\% (9.470\%) CTR, and 6.091\% (5.666\%) NFN on featured (single-columned) pages, which is a significant improvement on the Kuaishou live streaming platforms. These results indicate the significance of timeliness in live recommender systems and directly reducing the window size of the data streaming can result in substantial improvements.

Table ~\ref{tab:up2} 
shows the results of a four-day online A/B test after upgrading the five-minute data stream to our proposed Sliver data stream and adopting our time-sensitive re-reco strategy to reduce the time delay between request and impression. From this table, we can observe that our approach achieves further average improvements
of 8.304\% (6.765\%) CTR, and 3.697\% (2.783\%) NFN on two pages. This result validates the effectiveness of further enhancing the timeliness of live recommender systems while simultaneously considering label accuracy.

It's worth noting that each data stream upgrade results in large performance gains which further illustrates the importance of improving the timeliness of live recommendations from a data perspective.

\subsection{ System Deployment}

In Figure ~\ref{fig:service}, we show the architecture of the recommender system on Kuaishou live streaming platform. It is divided into two parts: online service and data stream loop. For the online service, the workflow is illustrated in the following:
\begin{itemize}[leftmargin=*]
\item  When a user launches Kuaishou App, the client first requests the live recommendation service which sequentially requests recall, pre-rank, and rank service to get the recommended live result.
\item The recommended live result will lack timeliness due to it may be not shown to the user in time. To address this problem, as discussed in section~\ref{sec:2}, we propose a re-reco strategy that determines whether the rank model needs to be re-requested based on whether the recommended live results are real shown or not. In this phase, the input candidate list is directly fetched from Redis which is stored in the first request phase to save resources of recall and pre-rank service.    
\end{itemize}
The workflow of the data stream, where our Sliver
approach is implemented and deployed, is described in the
following:
\begin{itemize}[leftmargin=*]
\item  After requesting the recommendation service to get the recommendation results, the data stream gets user features based on user info, the anchor and live  features from the kv store base on live index at the time of the request.
\item As for the label, the action log collects user behaviors after the impression and then is saved in the label store.  We adopt a 30s trigger to keep joining samples. If the behavior occurs, the trigger will do sample join.
\item The joined samples will incrementally update the rank model in the form of streaming learning.
\end{itemize}

\section{Conclusion}
In this paper, we try to address the problem of how to ensure both the timeliness and label accuracy in live streaming recommendation tasks. We propose a novel sliding window data stream paradigm Sliver to address this problem.  To ensure timeliness, we shorten the observation window to thirty seconds. To ensure label accuracy, we slide the observation window and adopt the explicit negative feedback of exiting the live room to avoid delayed feedback. Moreover, we propose a time-sensitive re-reco strategy to ensure the timeliness of the whole recommendation service and features. The experimental results demonstrate the effectiveness of the Sliver on multi-task live 
streaming recommendation. Sliver is deployed in the Kuaishou live streaming platform and achieves 6.765\%-8.304\% CTR and  2.788\%-3.697\% NFN improvement on two pages. 

\bibliographystyle{ACM-Reference-Format}
\balance
\bibliography{ref}
\clearpage

\appendix
\section{APPENDIX}
\label{app:A}
\begin{figure*}[ht]
\centering
    \subfigure[Featured Page]{
		\begin{minipage}[b]{0.23\textwidth}
			\includegraphics[height=8cm]{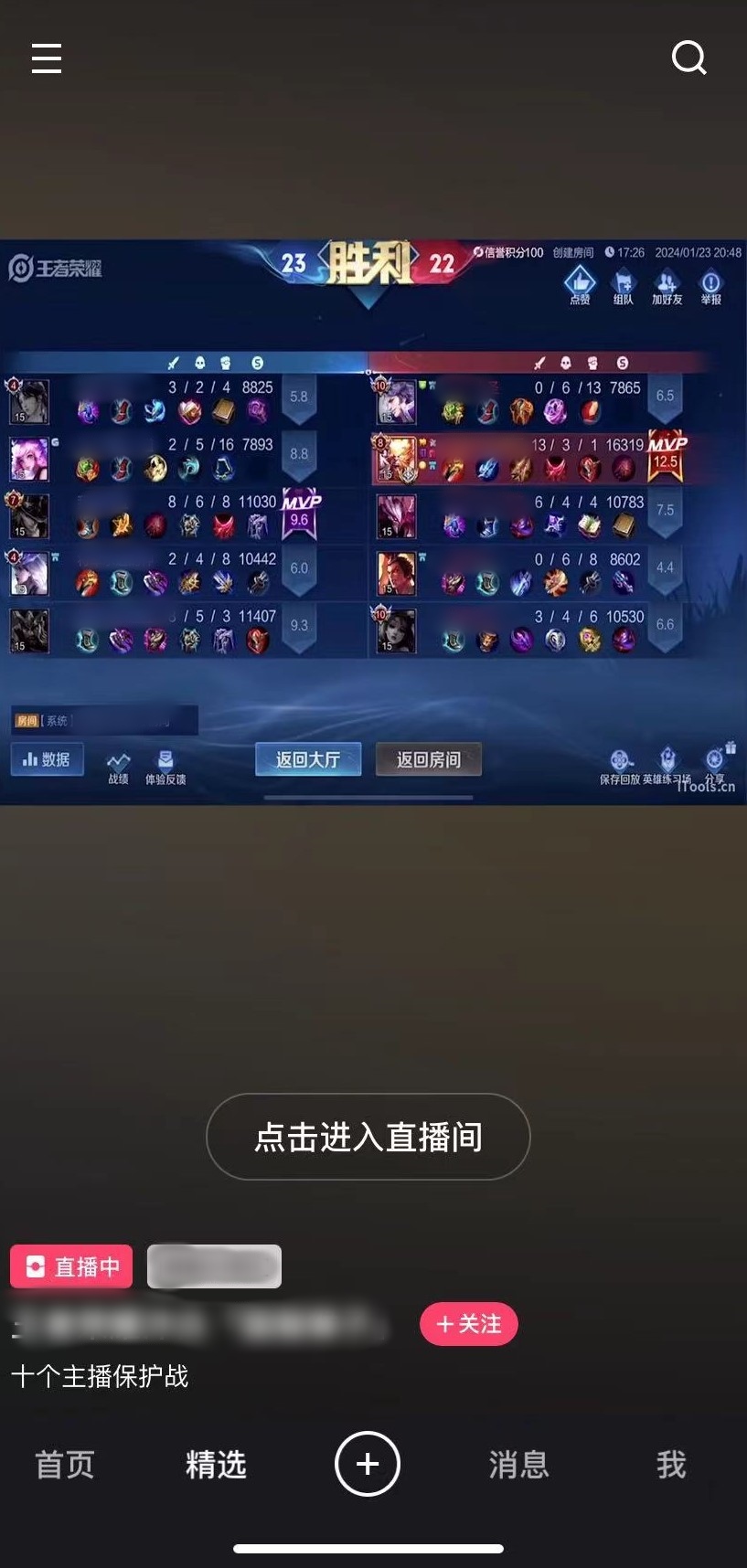}
		\end{minipage}
		\label{fig:jx}
	}
    \subfigure[Single-Columned Page]{
		\begin{minipage}[b]{0.23\textwidth}
			\includegraphics[height=8cm]{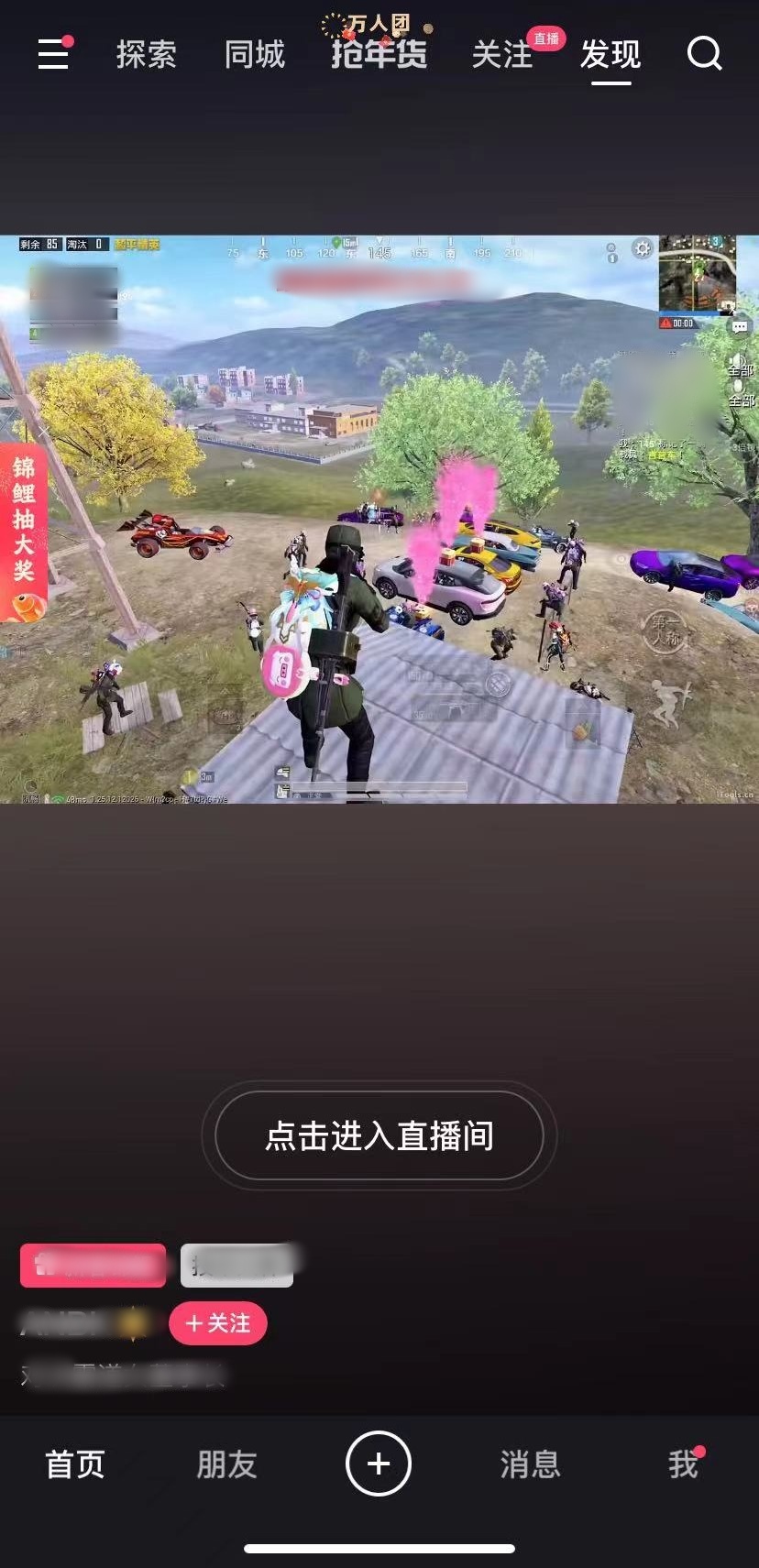}
		\end{minipage}
		\label{fig:js}
	}
 \subfigure[Live Room]{
		\begin{minipage}[b]{0.23\textwidth}
			\includegraphics[height=8cm]{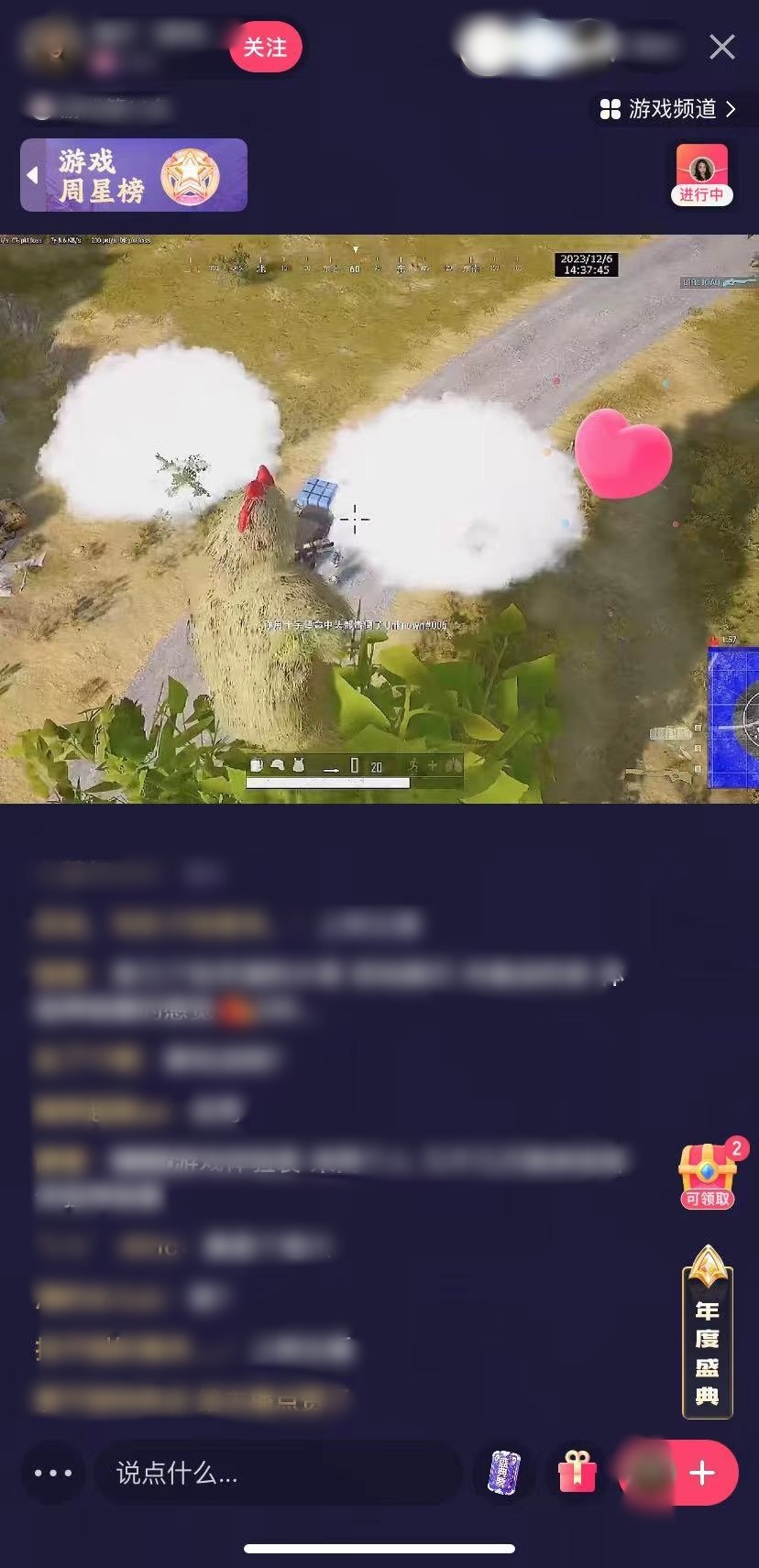}
		\end{minipage}
		\label{fig:like}
	}
    \subfigure[Exit]{
		\begin{minipage}[b]{0.23\textwidth}
			\includegraphics[height=8cm]{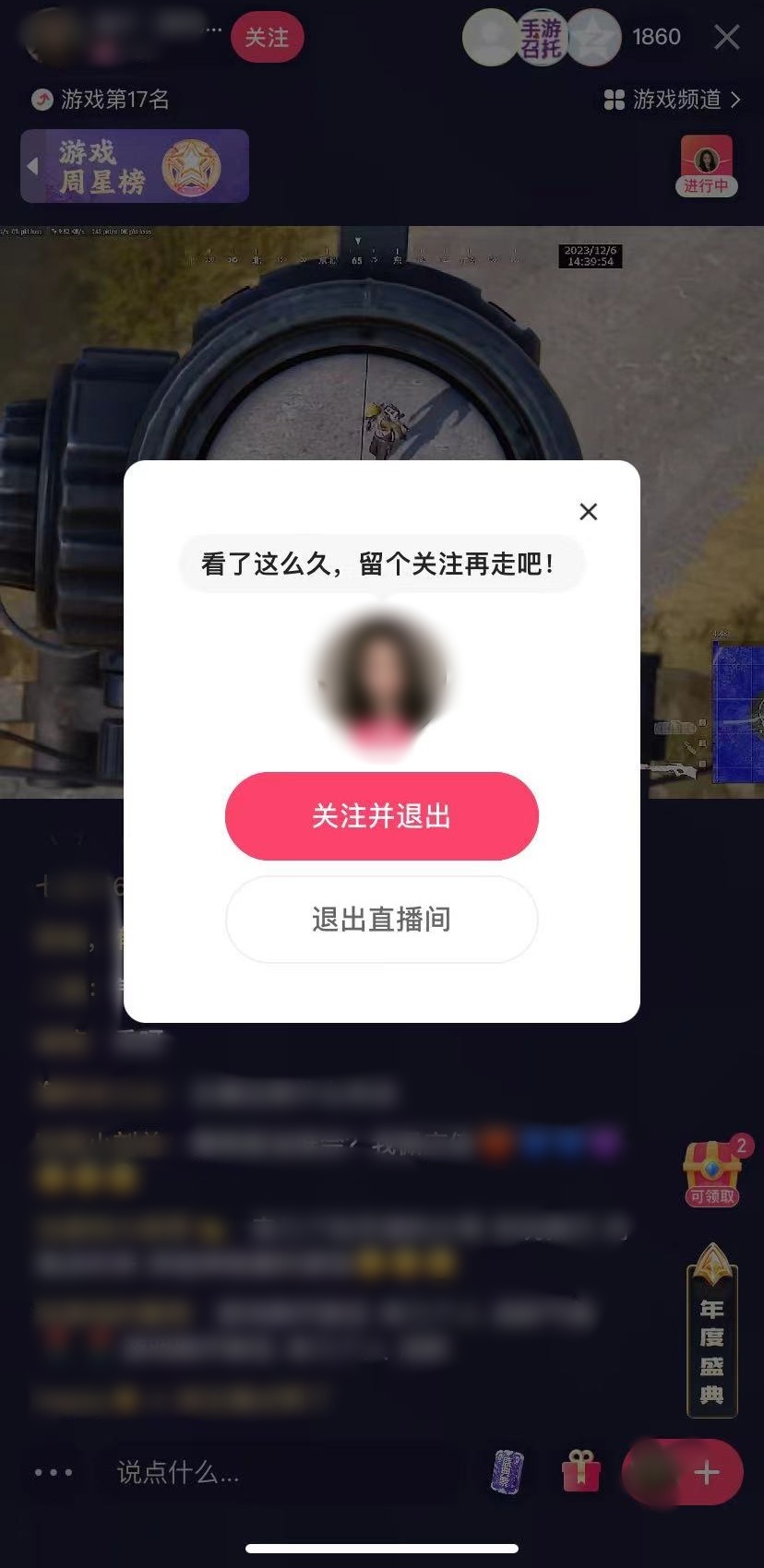}
		\end{minipage}
		\label{fig:exit}
	}
    \caption{This figure shows the live streaming recommendation scenario under two different pages in the Kuaishou APP. Our Sliver data stream is deployed under these two pages. }
    \label{fig:product}
\end{figure*}

\subsection{Live streaming Recommendation in Kuaishou APP}

Kuaishou APP is one of the largest short-video
and live-streaming platforms in China. Our Sliver data stream is deployed under two different pages: the featured page and the single-columned page. As shown in Figure ~\ref{fig:jx} and Figure ~\ref{fig:js}, the product form of these two pages is the up-and-down slide page. The results of live recommendations are ranked mixed with other types of items (e.g., short-video). After clicking into the live room, as shown in Figure~\ref{fig:like}, the audiences can do behavior such as like behavior (double click) and follow behavior. Note the follow behavior can also be performed before click behavior. If audiences don't want to watch this live room, they can exit the live room as shown in Figure~\ref{fig:exit}. Meanwhile, if the user does not click on the live stream and slide away, we also take it as an exit behavior.

\end{document}